\documentclass[11pt]{article}
\usepackage{authblk}

\usepackage[framemethod=TikZ]{mdframed}
\usepackage[linesnumbered,ruled,vlined]{algorithm2e}
\usepackage{pifont}
\usepackage[many]{tcolorbox}
\usepackage[utf8]{inputenc} 
\usepackage{algpseudocode}  
\usepackage{amsmath}
\usepackage{amssymb}
\usepackage{amsthm}
\usepackage{appendix}
\usepackage{blindtext} 
\usepackage{caption}
\usepackage{empheq}
\usepackage{geometry}
\usepackage{graphicx}
\usepackage{hyperref}  
\usepackage{cleveref}
\usepackage{soul}
\usepackage{subcaption}
\usepackage{tikz}
\usepackage{xcolor}
\usepackage[normalem]{ulem}

\usepackage{newunicodechar}
\hypersetup{
    colorlinks=true,
    linkcolor=blue,
    filecolor=blue,      
    urlcolor=cyan,
    citecolor = cyan,
}

\makeatletter  
\newif\if@restonecol  
\makeatother

\definecolor{dg}{RGB}{0,204,0}
\definecolor{dr}{RGB}{88,0,0}
\definecolor{db}{RGB}{0,0,88}

\newcommand{\darkgreen}[1]{{\color{dg} #1}}
\newcommand{\darkred}[1]{{\color{dr} #1}}

\newcommand{\ie}{\textit{i.e. }}

\newcommand{\bfy}{\boldsymbol{y}}
\newcommand{\bfx}{\boldsymbol{x}}

\newcommand{\cL}{\mathcal{L}}%

\newcommand{\cX}{\mathcal{X}}
\newcommand{\cY}{\mathcal{Y}}

\newcommand{\bR}{\mathbb{R}}
\newcommand{\cmark}{\darkgreen{\ding{51}}}%
\newcommand{\xmark}{\darkred{\ding{55}}}%

\newtcolorbox{mybox}[3][]
{colback=white,
colframe=#2!20,
colbacktitle=#2!20!,
coltitle=#2!20!black,
title={#3},
#1
}

\newtheorem{proposition-definition}{Proposition-definition}

\usepackage{listofitems} 
\usetikzlibrary{arrows.meta} 
\usepackage[outline]{contour} 
\contourlength{1.4pt}

\tikzset{>=latex} 
\colorlet{myred}{red!80!black}
\colorlet{myblue}{blue!80!black}
\colorlet{mygreen}{green!60!black}
\colorlet{myorange}{orange!70!red!60!black}
\colorlet{mydarkred}{red!30!black}
\colorlet{mydarkblue}{blue!40!black}
\colorlet{mydarkgreen}{green!30!black}

\tikzstyle{node xl}=[thick,circle,draw=myblue,minimum size=30,inner sep=0.5,outer sep=0.6]
\tikzstyle{node}=[thick,circle,draw=myblue,minimum size=20,inner sep=0.5,outer sep=0.6]
\tikzstyle{func} = [rectangle,draw=myblue, minimum size=26,inner sep=0.5,outer sep=0.6, fill=myblue!20!white]
\tikzstyle{node in}=[node,green!20!black,draw=mygreen]
\tikzstyle{node hidden}=[node,blue!20!black,draw=myblue]
\tikzstyle{node convol}=[node,orange!20!black,draw=myorange!30!black,]
\tikzstyle{node out}=[node,red!20!black,draw=myred]

\tikzstyle{node in xl}=[node xl,green!20!black,draw=mygreen]
\tikzstyle{node hidden xl}=[node xl,blue!20!black,draw=myblue]
\tikzstyle{node out xl}=[node xl,red!20!black,draw=myred]

\tikzstyle{connect}=[thick,mydarkblue] 
\tikzstyle{connect arrow}=[-{Latex[length=4,width=3.5]},thick,mydarkblue,shorten <=0.5,shorten >=1]
\tikzset{ 
  node 1/.style={node in},
  node 2/.style={node hidden},
  node 3/.style={node out},
  node 1xl/.style={node in xl},
  node 2xl/.style={node hidden xl},
  node 3xl/.style={node out xl},
  Dotted/.style={
    dash pattern=on 0.1\pgflinewidth off #1\pgflinewidth,line cap=round,
    shorten >=#1\pgflinewidth/2,shorten <=#1\pgflinewidth/2},
  Dotted/.default=3
}
\usetikzlibrary{
  arrows.meta, 
  fit, 
  positioning, 
  shapes, arrows
}

\AtBeginEnvironment{subappendices}{%
\chapter*{Appendix}
\addcontentsline{toc}{section}{Appendices}
\counterwithin{figure}{section}
\counterwithin{table}{section}
}

\usepackage{array}
\newcommand{\PreserveBackslash}[1]{\let\temp=\\#1\let\\=\temp}
\newcolumntype{C}[1]{>{\PreserveBackslash\centering}p{#1}}

\providecommand{\keywords}[1]
{
  \small	
  \textbf{\textit{Keywords---}} #1
}
\usepackage{caption}
\author[1]{Ruihua RUAN\thanks{Corresponding author.
Email: ruan@ceremade.dauphine.fr}}
\author[1]{Emmanuel BACRY}
\author[2]{Jean-François MUZY}
\affil[1]{CEREMADE, CNRS-UMR 7534, Universit\'e Paris-Dauphine PSL\protect\\
Place du Mar\'echal de Lattre de Tassigny, 75016 Paris, France}
\affil[2]{SPE CNRS-UMR 6134, Universit\'e de Corse
BP 52, 20250 Corte, France}

\title{Liquidity takers behavior representation  through a contrastive learning approach}
\begin{document}
\maketitle
\begin{abstract}
    Thanks to the access to the labeled orders on the CAC40 data from Euronext, we are able to analyze agents' behaviors in the market based on their placed orders. In this study, we construct a self-supervised learning model using triplet loss to effectively learn the representation of agent market orders. By acquiring this learned representation, various downstream tasks become feasible. In this work, we utilize the K-means clustering algorithm on the learned representation vectors of agent orders to identify distinct behavior types within each cluster.
\end{abstract}
\keywords{agent behaviors, agent-based model, contrastive learning, triplet loss, clustering}

\section{Introduction}

Deep learning has achieved great success in recent years, mainly due to advances in machine learning algorithms and computer hardware. As a result, it has become an indispensable tool in a wide range of fields, both in research and in practical applications. Specifically, in finance, deep learning has been applied extensively to predict stock prices movements using limit order book data. This technique is particularly effective in handling complex data which statistical models often struggle to manage. Notable works in the recent literature include \cite{zhang2019deeplob, sirignano2019deep, sirignano2019universal, zhang2021deep}. 

In particular, contrastive learning (CL) is a powerful technique in deep learning that has led to significant advances in representation learning. It has been widely applied, especially in vision domain, as demonstrated by the success of works such as \cite{chen2020simple, grill2020bootstrap, he2020momentum}. In the domain of time-series analysis, CL has also shown great potential. For example, Contrastive Predictive Coding (CPC) of \cite{oord2018representation} employed a latent space to capture historical information and predict future observations, and has demonstrated impressive results in speech recognition tasks. In healthcare, authors in \cite{mohsenvand2020contrastive} applied CL on electroencephalogram data while \cite{mehari2022self} used it to electrocardiography data. In finance, CL has been used for stock trend prediction \cite{hou2021stock}, and financial time series forecasting \cite{wu2020conditional}. 

In financial market, the collective actions of agents on the limit order book determines the macroscopic evolution of the market. Therefore, to fully understand the dynamics of a market, it is crucial to comprehend the roles and strategies of individual agents. However, due to the challenge of accessing confidential trading data, only a few studies have been conducted in the area of characterizing market participants. For example, Brogaar and others have studied limit order book data with agents labeled as either High-frequency traders (HFTs) or Market makers (MMs) in \cite{brogaard2010high, brogaard2014high}. The authors of \cite{hagstromer2013diversity} have analyzed the different behaviors of HFTs and MMs. With access to agent identities, \cite{kirilenko2017flash} classified the agents into HFTs, MMs, fundamental buyers, fundamental sellers and opportunistic traders, and studied their behavior before and after the flash crash of may 2010. A recent research in this area was achieved by Cont et al. in their work \cite{cont2023analysis}, where they analyzed limit order book data from the broker view and grouped the agents into four groups, for each they detailed descriptions of the properties. A study that deserves special attention in our research is the work \cite{cartea2023statistical}. In their study, the authors presented statistical models designed to predict the behavior of trading algorithms using data from Euronext Amsterdam. By extracting the coefficients from their prediction model, they identified three distinct categories of trading algorithms prevalent in the market: directional trading, opportunistic trading, and market making. 

In this paper, our objective is to analyze and characterize the different behaviors of the agents. More specifically, we will study successions of any fifty consecutive market orders placed by any agent. Each order is defined by eight features (see Section \ref{sec:data}), resulting in a sample matrix of size $50 \times 8$. The inner structure of such a sample can be complex and challenging to represent using classical methods. To address this challenge, we aim to learn a representation (i.e., embeddings) that can effectively embed each such sample into a lower-dimensional vector space. 
We propose to use a self-supervised contrastive learning approach using a triplet loss \cite{schroff2015facenet}. For each "anchor" sample from an agent two other samples (not overlapping in time) are chosen : 
\begin{itemize}
    \item[-] one (positive) sample from the same agent
    \item[-] one (negative) sample from another agent.
\end{itemize}
The pretext task, from which the embeddings are learned, consists in trying to identify the positive sample (through the use of the triplet loss). All the sample are taken over a two hours period of time during the same day, so that the positive sample and the reference sample, corresponding from the same agent and being close in time, can be considered hopefully as corresponding to the "same" structure/strategy.

The learned embeddings can be utilized for various downstream tasks, such as clustering and classification. In this work, we will apply the K-means clustering algorithm to the learned embeddings, by doing so, we aim to reveal the different strategies employed by agents and the development of their strategies over time. To the best of our knowledge, we are the first to propose a contrastive learning method on limit order book representation.

\paragraph{Outline.} 
This paper is organized as follows. In Section \ref{sec:prelim}, we recall some important concepts and terminology, including limit order book, liquidity makers and takers, and contrastive learning with triplet loss. We will also properly formulate the main problem to be addressed in this paper. In Section \ref{sec:data}, we describe the data used as well as the preprocessing steps. Section \ref{sec:nn} represents the neural network architectures, implementation information and evaluation metrics. We demonstrate the importance of some features. In Section \ref{sec:cluster}, we apply the K-means clustering algorithm, a downstream task, to the learned embeddings. We analyze the properties of each cluster and the clustering results of each agents. Finally, concluding remarks and discussion are provided in Section \ref{sec:conclusion}.

\section{Preliminaries}\label{sec:prelim}
In order to provide a comprehensive understanding of the concepts and terminology used in this paper, we will begin by briefly reviewing several important concepts. These include concepts in financial markets background, such as limit order books, liquidity takers, as well as a loss function for contrastive learning, namely the triplet loss. At the end of this section, we will proceed to formulate the main problem to be addressed in this study.

\subsection{Limit order book}
A \textbf{limit order book} (LOB) is an auction mechanism used in financial markets to record the buy or sell orders placed by traders. These orders can be categorized into three major types : limit orders, cancellation orders, and market orders. 

A \textbf{market order} is an order to buy or sell a stock at the market's current best available price, which typically ensures an immediate execution. Conversely, a limit order is a to buy or sell order at a specific price, which cannot be executed immediately. This is because the current market quotes do not match the trader's desired target price. In this case, the limit order will join the queue in the limit order book and wait until it can be executed at the desired price or a better one, unless it has been canceled. The action to cancel a limit order corresponds to a cancellation order, which removes an unfilled order from the queue. We refer the interested readers to the reference \cite{gould2013limit} for a very nice review of limit order book concepts.

Furthermore, it is important to provide definitions for \textbf{aggressive trades} and \textbf{passive trades}. When agent $\alpha$ places a market order, it effectively can be seen as a match of two orders. They are respectively an existing limit order at price $p$ in the queue placed by agent $\beta$, and a marketable limit order placed by agent $\alpha$ that matches this price $p$. This market order can be seen as both an aggressive trade for agent $\alpha$ and a passive trade for agent $\beta$. It is worth noting that these terminologies may differ from other definitions that one can find elsewhere.

\subsection{Liquidity takers \textit{vs.} Liquidity providers}
In financial markets, participants can be broadly classified into two categories : liquidity providers and liquidity takers. Liquidity providers, also known as market markers, are the agents who place limit orders on both sides of the market (buy and sell) and attempt to earn the bid-ask spread. Conversely, \textbf{liquidity takers}, typically traders and investors, seek to earn profits from the price movement of asset or use the price movement as a hedge to the other positions in their portfolio. In traditional markets, market makers are usually designated by the market while in modern markets, anyone can be a market maker. In fact the distinction between liquidity providers and takers is not clear-cut. 

In this paper, we will focus on the the behavior of liquidity takers through the analysis of their aggressive trades in the LOB. Later in this paper, we will demonstrate that some highly active liquidity takers are also significant liquidity providers. For instance, Member 11, 12 and 24 in Fig. \ref{fig:passive_ratio}) can be considered as such agents.  

\subsection{Self-supervised learning with Triplet loss}
Self-supervised learning (SSL) is a machine learning approach, which processes unlabeled data to obtain useful representations that are helpful for various downstream tasks. Among self-supervised method, contrastive learning is a very popular technique, notably used for computer vision tasks (for instance SimCLP \cite{chen2020simple}, BYOL \cite{grill2020bootstrap}, MoCo \cite{he2020momentum}, Barlow twins \cite{zbontar2021barlow}). Its aim is to learn a representation function that embeds similar inputs close together and dissimilar inputs far apart. Over the years, the loss functions used in contrastive learning have evolved from a simple comparison between one positive and one negative sample \cite{chopra2005learning} to  multiple positive and negative samples \cite{sohn2016improved, gutmann2010noise, oord2018representation}. 

In this work, we apply our deep neural networks, which are equipped with the triplet loss, to learn a representation function for limit order book data. The Triplet Loss was first introduced by Schroff et al. in their 2015 paper \cite{schroff2015facenet}, where it was used for face recognition of individuals under varying poses and angles. Since then, it has become a widely used loss function for supervised similarity tasks. As illustrated by Fig.\ref{fig:triplet}, the fundamental idea behind the Triplet Loss is to learn a representation function $f(\cdot)$ that brings inputs that match (referred to as positive inputs) closer to the reference input (referred to as the anchor) and pushes away inputs that do not match (referred to as negative inputs). The triplet loss function can be defined as following : 
\begin{equation}\label{eq:triplet}\tag{$\mathcal{L}_1$}
    \cL_{triplet} = \sum_{i=1}^N \max\Big(||f(X_i^a)-f(X_i^p)||_2^2 - ||f(X_i^a)-f(X_i^n)||_2^2 + \gamma, 0\Big)
\end{equation}
where 
\begin{itemize}
    \item[-] $f:\cX \to \bR^d$ is a representation function that embeds an input to a d-dimensional Euclidean space $\bR^d$.
    \item[-] $\gamma$ is a margin between positive and negative pairs, the margin value is added to push negative samples far away.
    \item[-] $X^a$ indicates \textbf{Anchor} sample, $X^p$ indicates \textbf{Positive} sample, $X^n$ indicates \textbf{Negative} sample.
\end{itemize}
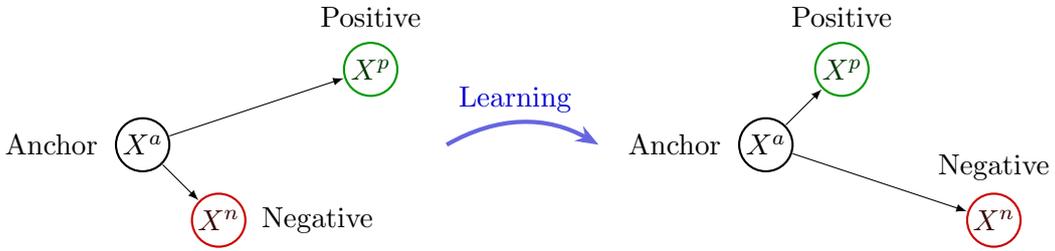
\begin{figure}[h]
    \centering
\begin{tikzpicture}
  \node[node 1, black] at (-5,0) (xa) {$X^a$};
  \node[node 3] at (-4,-1) (xn) {$X^n$};
  \node[node 1] at (-2,1) (xp) {$X^p$};

  \node[node 1, black] at (3.2,0) (xa1) {$X^a$};
  \node[node 3] at (6.2,-1) (xn1) {$X^n$};
  \node[node 1] at (4.2,1) (xp1) {$X^p$};

  \draw [draw=myblue!60, ultra thick, arrows=-{Stealth[length=3mm]}] (-1,0) to [bend left] (1,0);
  \draw [->] (xa)--(xn);
  \draw [->] (xa)--(xp);
  \draw [->] (xa1)--(xp1);
  \draw [->] (xa1)--(xn1);

  \node[black] at (-6.2,0) {Anchor};
  \node[black] at (2,0) {Anchor};
  \node[black] at (-2,1.7) {Positive};
  \node[black] at (4.2, 1.7) {Positive};
  \node[black] at (-2.7, -1) {Negative};
  \node[black] at (6.2, -.3) {Negative};

  \node[myblue] at (-0.1, 0.6) {Learning};

\end{tikzpicture}
    \caption[Triplet Illustration]{Triplet Loss illustration. The Triplet Loss minimizes the distance between the anchor $X^a$ and the positive $X^p$, and maximizes the distance between the Anchor $X^a$ and the negative $X^n$.}
    \label{fig:triplet}
\end{figure}

\subsection{Problem formulation}
\begin{figure}[ht]
    \centering
    \includegraphics[width=0.8\linewidth]{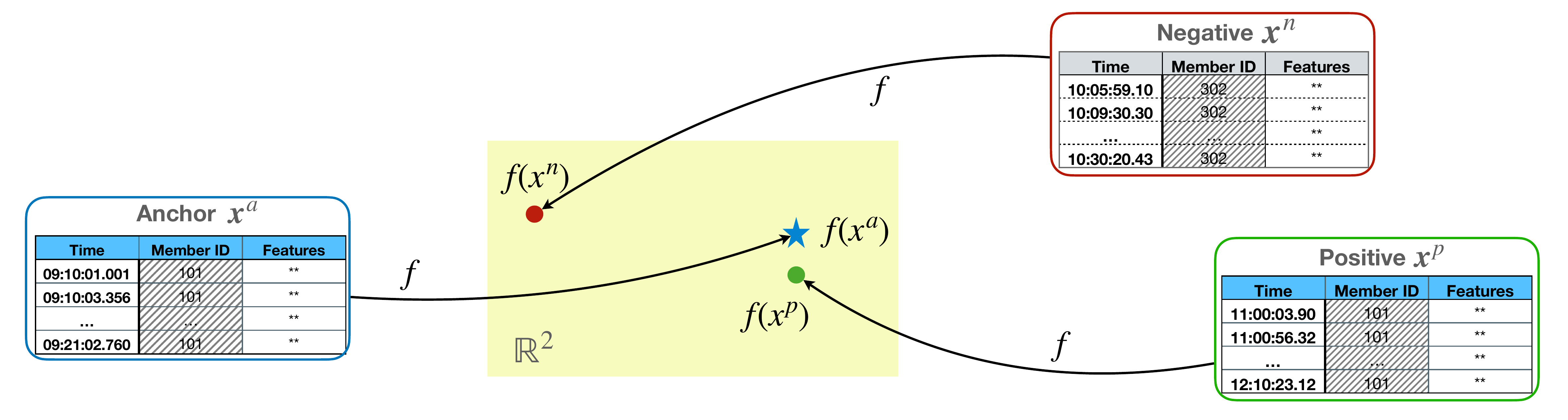}
     \caption{Illustration of model learning. $X$ is a sequence of consecutive market orders and $f(X)$ is its vector representation in $\bR^2$. The triplet loss minimizes the distance between the samples from the same agent $||f(X^p)-f(X^a)||_2$ and maximize the distance between the samples from different agents $||f(X^n)-f(X^a)||_2$.}
    \label{fig:encoder}
\end{figure}

The objective of this work is to develop a robust method for representing a sequence of consecutive market orders sent by the same agent. To this end, we introduce a novel approach that employs a deep neural network with an LSTM architecture, equipped with a triplet loss function. 
Fig. \ref{fig:encoder} gives an illustration of an example framework (when $d=2$). 

Since a sequence of orders is highly structural, determining the similarity between two sequences of orders can be very challenging. However the $\bR^d$ space is widely recognized and offers a straightforward measure of distance between two points. As a result, the representation function $f$ establishes a connection between the intricate order book space and a more comprehensible $\bR^d$ space. 

Once a comparison between two sequences of orders becomes possible, one can apply various downstream tasks. In this work, our focus lies in grouping these sequences of orders from different agents, represented by their images in $\bR^d$,  to several clusters. Through this clustering process, we expect to uncover the trading behavior and strategy of these agents. 

\section{Data Description}\label{sec:data}
In this present work, we analyze the limit order book (LOB) of the front month\footnote{The term "front month" refers to the nearest expiration date in futures trading.} CAC40 index future contracts. The data was obtained from the Euronext market and spans a period of 300 consecutive trading days, from January 6th, 2016 to March 7th, 2017, between 9:00 am and 5:00 pm each day. Let us again mention that this task focuses only on market orders in the LOB rather than all types of orders. 

We present a network that utilizes the Triplet Loss \eqref{eq:triplet} and takes consecutive market orders of an agent as inputs. Through training this network, our goal is to obtain a robust representation function that maps order book inputs to a lower-dimensional vector space. Similarity between two order book inputs is determined based on whether they belong to the same agent or not.

\subsection*{Agents selection}
Out of our 300-day dataset, we have identified 170 active Members. Despite this, the majority of these members either have limited daily order volume or show only brief periods of activity. In this study, we only consider Members who have placed at least 200 market orders each day on more than 45 separate trading days. This selection process results in a pool of 30 highly "market order" active Members. Remarkably, each of these selected agents has placed no less than 15,000 market orders during the designated period. 

Once more, we only consider the aggressive trades (market orders) executed by an agent during a period and do not include the passive trades, which are executed by market orders placed by another agent that fully or partially match the limit orders of this agent. However, it's important to note that passive trades can have a significant impact on an agent's behavior, and we plan to address this issue in our future work.

In order to have a deep insight into the 30 members that we selected and measure how much these 30 members are weighted in the market, we have conducted the following statistics.

\begin{enumerate}
    \item Let us define the "actions at L1" as all the orders (limit, cancellation and market) which are executed at the best bid or best ask price levels. The previous 30 selected members (which place the most market orders) are in fact among the top 40 most active agents at L1, with the top 20 most active agents included in this group. In simpler terms, these 30 selected members can be considered the most influential agents at L1. 
    
    \item We also provide the ratio of the number of passive trades to the number of aggressive trades for each agent in Fig \ref{fig:passive_ratio}, in order to gain a better understanding on the visible  of these agents. A high passive-aggressive ratio indicates a more market-marker-like agent, who provides more liquidity than takes liquidity from the market. 
    \begin{figure}[ht]
        \centering
        \includegraphics[width=\linewidth]{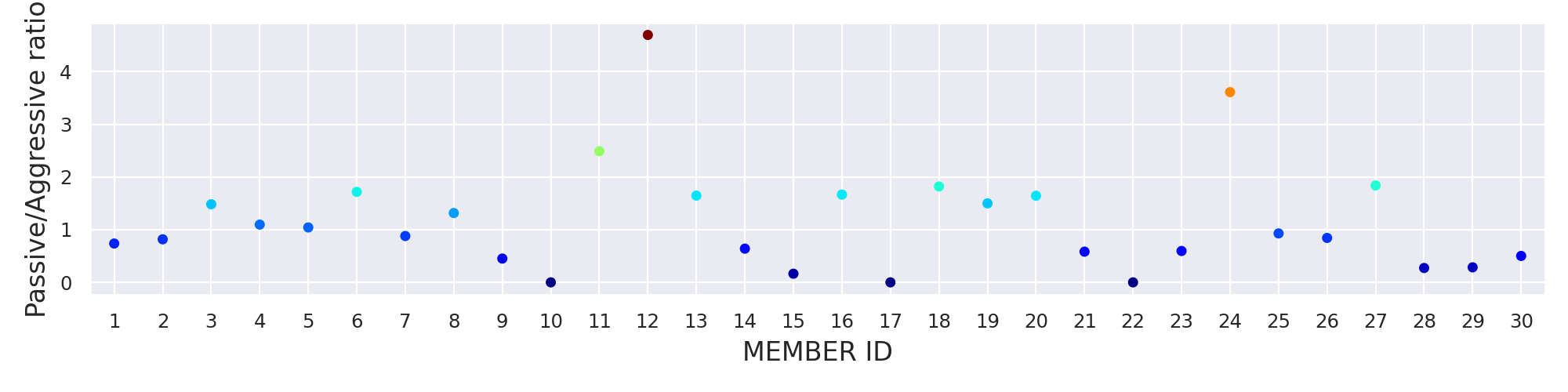}
        \caption{The ratio of passive trades to aggressive trades for each Member}
        \label{fig:passive_ratio}
    \end{figure}
\end{enumerate}

These statistics demonstrate that by focusing on those who place significantly aggressive trades, we have effectively taken into account the majority of the most important active market participants.

\subsection*{Order features and Input data} 
In this work, we have chosen to describe a market order $\bfx$ by the following features,
\begin{itemize}
    \item[-] $t$ (timestamps) : the timestamp which records the time point when an order is executed
    \item[-] $q_T$ (Quantity)  : the amount of stocks traded (this value is always equal to or less than the size proposed by the trader)
    \item[-] $s$ (Side) : whether an order is a buy or sell order
    \item[-] $M$ (Limit to trade modification):  the specific type of action, $M=1$ represents the modification of an existing limit order to make it aggressive, while $M=0$ signifies the placement of an aggressive order that is immediately executed
    \item[-] $P^b_1(t-)$ : the best bid price immediately before the execution of this order
    \item[-] $P^a_1(t-)$ : the best ask price immediately before the execution of this order
    \item[-] $Q^b_1(t-)$ :  the volume of limit orders at the best bid (best bid queue size)
    \item[-] $Q^a_1(t-)$ :  the volume of limit orders at the best ask (best ask queue size)
\end{itemize}

It results that an order is represented as an 8-dimensional vector.
Each input $X$ is a sequence of 50 consecutive market orders executed by one agent $\alpha$, $X=(\bfx_i)_{i=1,2,...,50}$, corresponding to a matrix in $\bR^{50\times 8}$. The set of inputs labeled by $\alpha$ is denoted by $\mathcal{Y}_\alpha$.
In the Triplet Loss approach, a single input is composed of $X^a, X^p, X^n$, where $X^a$ and $X^p$ (note that $X^a \neq X^p$) come from the same agent $\alpha$, while $X^n$ is sourced from a different agent $\beta$. In other words,  $X^a$ and $X^p$ belong to the set $\cY_\alpha$, while $X^n$ belongs to another set $\cY_\beta$, with $\beta\neq\alpha$. The triplets are constructed locally in time, the positive sample $X^p$ and the negative sample $X^n$ are required to be "temporally close" to the anchor sample. In this work, two samples are considered "temporally close" if the time interval between their first order's timestamps is less than 2 hours.

The use of a local model in this work is motivated by the dynamic nature of agent behavior. As the goal is to learn the representation of sequences of orders of an agent, it is expected that an agent's strategy may change over time. In such a scenario, it would not be appropriate to force inputs that are far apart in time to match, even though they are from the same agent. By utilizing a local model, the contrastive learning approach is leveraged to better capture the dynamic nature of agent behavior.

\section{Implementation details and numerical results}\label{sec:nn}
In this study, we employ Long Short-Term Memory (or simply LSTM) to process the sequences of market orders. LSTM is a variation of of RNN which was designed to address the problem of vanishing gradients in standard RNNs \cite{hochreiter1997long}. Compared to another variant Gated recurrent unit (GRU), LSTM has a more complex structure. It includes memory cells, input gates, forget gates and output gates. Specifically, in this work, we applied stacked LSTMs, which are well-known for their ability to handle more complex models and deliver improved performance compared to the simple LSTM architecture \cite{sutskever2014sequence}. 
\subsection{Inputs and Hyperparameters}
We conducted tests on multiple sets of input sample features, and we present three of these sets in Table \ref{tab:feat}. In next subsection, we will introduce an evaluation metric and demonstrate that the best set of features is the (Basic+M+QS). Additionally, we performed tests with an extended set of features, including the queue sizes at level 2 and 3 in addition to the eight features, however we found that they do not exert a significant impact on the current task. As a result, we conclude that these eight features listed in Table \ref{tab:feat} are sufficient for our task.
\begin{table}[ht]
    \centering
    \begin{tabular}{cccc}
        \hline
        Features & Basic & Basic+M & Basic+M+QS \\
        \hline
        Time (interevent time) & \cmark & \cmark & \cmark\\
        Quantity  & \cmark & \cmark & \cmark \\
        Side (buy or sell) &  \cmark & \cmark & \cmark \\
        Limit to trade modification &  \xmark & \cmark & \cmark \\
        Best bid price &  \cmark & \cmark & \cmark \\
        Best ask price &  \cmark & \cmark & \cmark \\
        Best bid qty & \xmark & \xmark & \cmark \\
        Best ask qty &  \xmark & \xmark & \cmark \\
        \hline
    \end{tabular}
    \caption{3 types of input : Basic, Basic+M, Basic+M+QS. "M" stands for limit to trade modification and "QS" stands for the best level queue sizes.}
    \label{tab:feat}
\end{table}

\begin{figure}[ht]
    \centering
    \includegraphics[width=0.7\linewidth]{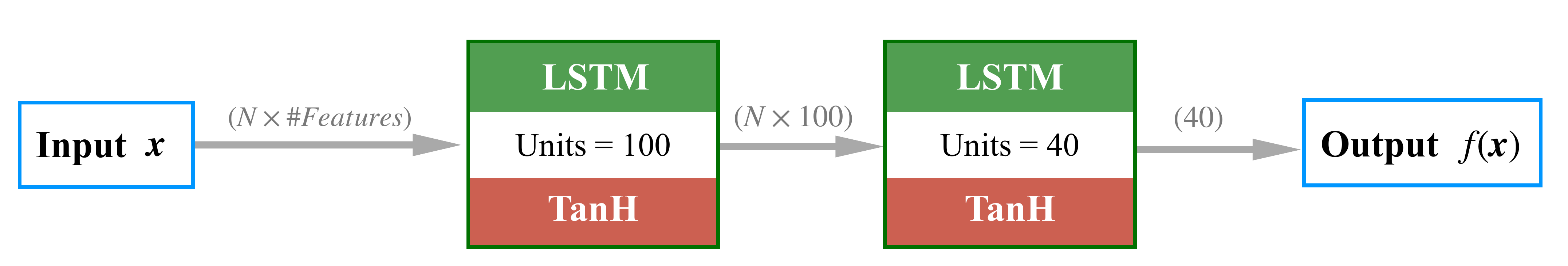}
    \caption{Encoding model architecture schema for one sample.}
    \label{fig:triplet_schema}
\end{figure}

The LSTM network used in this work has a stacked architecture with two hidden layers. The first layer consists of 100 units, while the second layer has 40 units. The encoded representation of the input sequence is obtained from the last output of the second layer, \ie the dimension of the embedding space is $d=40$ (We tried also other smaller output dimensions but they did not perform as well). The margin in triplet loss $\gamma$ was set to 0.5. See Fig \ref{fig:triplet_schema} for a model architecture illustration.
\begin{equation}\label{eq:nn}\tag{$NN1$}
    \begin{split}
        \text{\textbf{Input} : } & \text{triplet } (X^a, X^p, X^n)\\
        \text{\textbf{Encoder} : } & f = f^2 \circ f^1 \text{ with }
        \begin{cases}
            f^1(\cdot) = LSTM (p \to 100)\\
            f^2(\cdot) = LSTM (100 \to 40)
        \end{cases}\\
        \text{\textbf{Output} : } & (f(X^a), f(X^p), f(X^n)) \\
        \text{\textbf{Loss} : } & \max\big(||f(X^a)-f(X^p)||_2^2 - ||f(X^a)-f(X^n)||_2^2 + \gamma, 0\big)
    \end{split}
\end{equation}

The PyTorch library \cite{paszke2019pytorch} was primarily used to implement the neural networks. To train the networks, we used the Adam optimizer \cite{Kingma2015} with a learning rate of 0.002 and a batch size of 64.  The training process was conducted on a single NVIDIA GPU Tesla P4. The training process was stopped after 500 epochs. 
And early-stopping was not implemented due to the the effectiveness of the triplet loss in preventing overfitting. 

\subsection{Numerical results}
To prevent any data leakage, the 300 consecutive trading days are divided into two distinct sets, namely training days and test days. The numbers of training days and of test days follow a ratio of 4:1.  Therefore the training inputs $\mathcal{T}_0$ and the test inputs $\mathcal{T}_1$ are extracted from 240 training days and the remaining 60 test days respectively. We introduce an evaluation metric for the test data $\mathcal{T}_1$, called the failure rate. This metric is defined by the following formula:
$$r = \cfrac{\mid\left\{{i\in \{1,2,...,N\}} \text{ such that } ||f(X_i^a)-f(X_i^n)||<||f(X_i^a)-f(X_i^p)||\right\} \mid}{N}$$ 
where $N=|\mathcal{T}_1|$, and the test data $\mathcal{T}_1$ is defined as $\mathcal{T}_1 = \bigl\{(X_i^a, X_i^p, X_i^n),i=1,2,...,N\bigr\}$.
More precisely, when considering a particular agent $\alpha$, the failure rate can be expressed as follows : 
$$r_\alpha = \cfrac{\mid\left\{{i\in \{1,2,...,N\}} \text{ such that } X_i^a\in\cY_\alpha \text{ and } ||f(X_i^a)-f(X_i^n)||<||f(X_i^a)-f(X_i^p)||\right\} \mid}{\mid\left\{{i\in \{1,2,...,N\}} \text{ such that } X_i^a\in\cY_\alpha\right\}\mid}$$ 
Here $\mid\cdot\mid$ stands for the cardinality of a set. The quantity $(1-r_\alpha)$ is the proportion of triplets where the positive and negative sample are correctly distinguished. 

\begin{figure}[h]
    \centering
    \includegraphics[width=\linewidth]{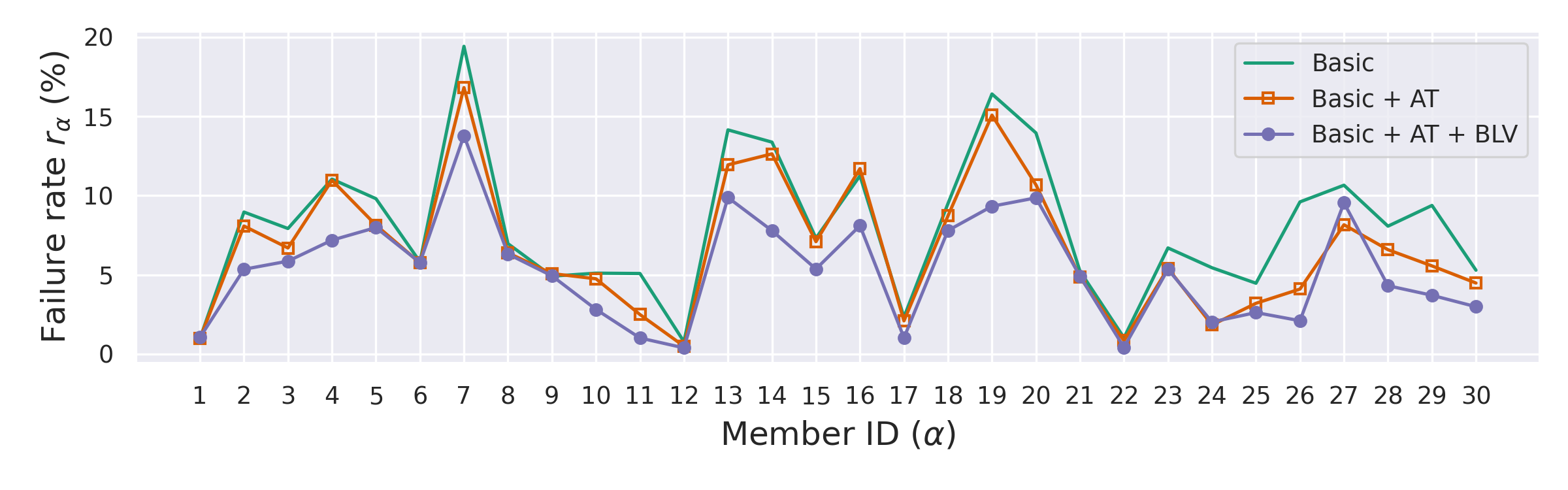}
    \caption{Failure rate for each agent in different scenarios. }
    \label{fig:fail_rate}
\end{figure}

Table \ref{tab:eval} gives the unconditional failure rates of the 3 feature sets. It reveals that the action type and the best bid/ask price level queue sizes play a crucial role in this task. The evaluation comparison, conditional on agents, for these feature sets is illustrated in Figure \ref{fig:fail_rate}. 

\begin{table}[ht]
    \centering
    \begin{tabular}{cccc}
        \hline
        Features & Basic & Basic+M & Basic+M+QS \\
        \hline
        Failure rate (r) & 8.03\% & 6.72\% & 5.32\%\\
        \hline
    \end{tabular}
    \caption{Evaluation results for the 3 types of input : Basic, Basic+M, Basic+M+QS.}
    \label{tab:eval}
\end{table}

\section{Downstream task : Clustering}\label{sec:cluster}
So far, we have acquired a learned representation function for market order sequences. In this section, we will test this representation in a downstream task that consists in performing a clustering of agent behavior. 

Cluster analysis is the task of grouping or segmenting a collection of objects into subsets or "clusters", such that the objects within the same cluster are more similar to each other than those assigned to different clusters (see 14.3 in \cite{hastie2009elements}). As detailed below, we will apply K-means cluster analysis to the encoded orderbook samples. To ensure a comprehensive analysis, we extract over 3000 samples for each agent from the market orders. Trough the neural network equipped with triplet loss, we obtain an encoder function, $f$, which can map a sequence of market orders to a lower dimensional vector that is interpretable. With the clustering of these lower dimensional vectors, we hope to uncover patterns that are not possible to discover through traditional statistical methods. More precisely, we aim to group the collection of trade orders by different agents into several subsets, and each subset will be considered as a trading strategy. Let us note that an agent can belong to multiple clusters and a cluster may encompass samples from different agents.

\subsection{K-mean clustering}
\begin{figure}[ht]
    \centering
    \includegraphics[width=\linewidth]{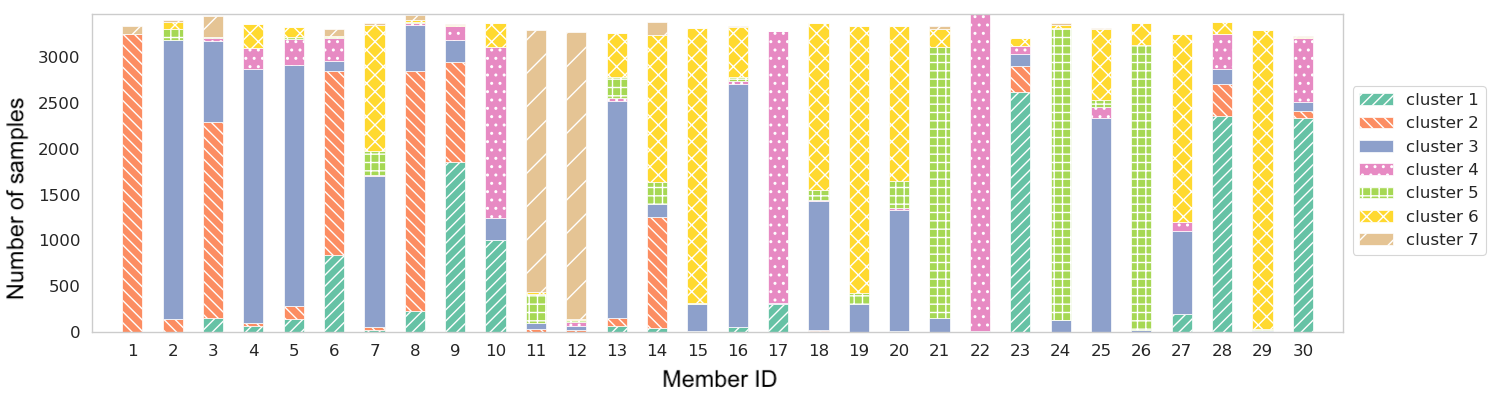}
    \caption{K-means clustering results. Each agent is represented by a vertical bar, which may consist of one or multiple segments. Each segment corresponds to the agent's samples assigned to a specific cluster.}
    \label{fig:kmeans}
\end{figure}

K-means clustering \cite{lloyd1982least, macqueen1967classification} is one of the most popular clustering methods. Given a set of observations $(\bfy_1, \bfy_2,...,\bfy_n) \in \bR^{n\times d}$, K-means clustering seeks to  minimize the within-cluster sum of squared deviations by assigning each observation $\bfy_j$ to its nearest cluster center. To formulate the task mathematically, K-means algorithm assigns these observations to $k^*$ clusters $\{Y_1, Y_2,...,Y_{k^*}\}$ by solving the following optimization problem. The cluster centers, denoted by $\mu_i$, are updated iteratively until convergence. 
$$
\min\sum_{i=1}^{k^*} \sum_{\bfy\in Y_i} \|\bfy - \mu_i\|^2
$$
In practice, in order to apply K-means, one must select the number of clusters $k^*$. In this work, we aim to apply K-means clustering to group the encoded orderbook samples into subsets of similar strategies. In order to provide a comprehensive view of strategies, it is desirable to have a relatively small  number of clusters compared to the total number of agents. A variety of methods have been proposed in the literature to determine the number of clusters $k^*$, such as the "Elbow" method, the gap statistic, the Silhouette method, etc. In this work, we employ the "Elbow" method to determine the optimal number of K-means clusters, which is set to 7. Figure \ref{fig:kmeans} displays the K-means clustering result.

One can remark that several Members belong to only one cluster, such as members  1,11,12,17 and 22. It is not surprising as these members have consistent and especially low failure rate (as shown in Figure \ref{fig:fail_rate}), showing that they are rather distinctive. On the other hand, many agents belong to several clusters which may be interpreted by the fact that their strategy is changing over time.
\subsection{Characterizing clusters by indicators}
In order to have a deeper understanding of these clusters and how they differ from one another, we will evaluate them using a set of indicators. To start, we present each indicator and plot the evaluation of input data for each cluster based on this indicator. These box plots show the quantile range (25th percentile, median and 75th percentile) for the inputs of each cluster. 
\paragraph{Frequency:} If $\delta t$ stands for the average interevent time (calculated as $\frac{1}{49}(t_{50}-t_1)$), the frequency indicator $\cfrac{60}{\delta t}$ represents the average number of trades per minute. The higher the value, the more frequently market orders are being placed by the agent.
\begin{figure}[h]
    \centering
    \includegraphics[width=0.5\linewidth]{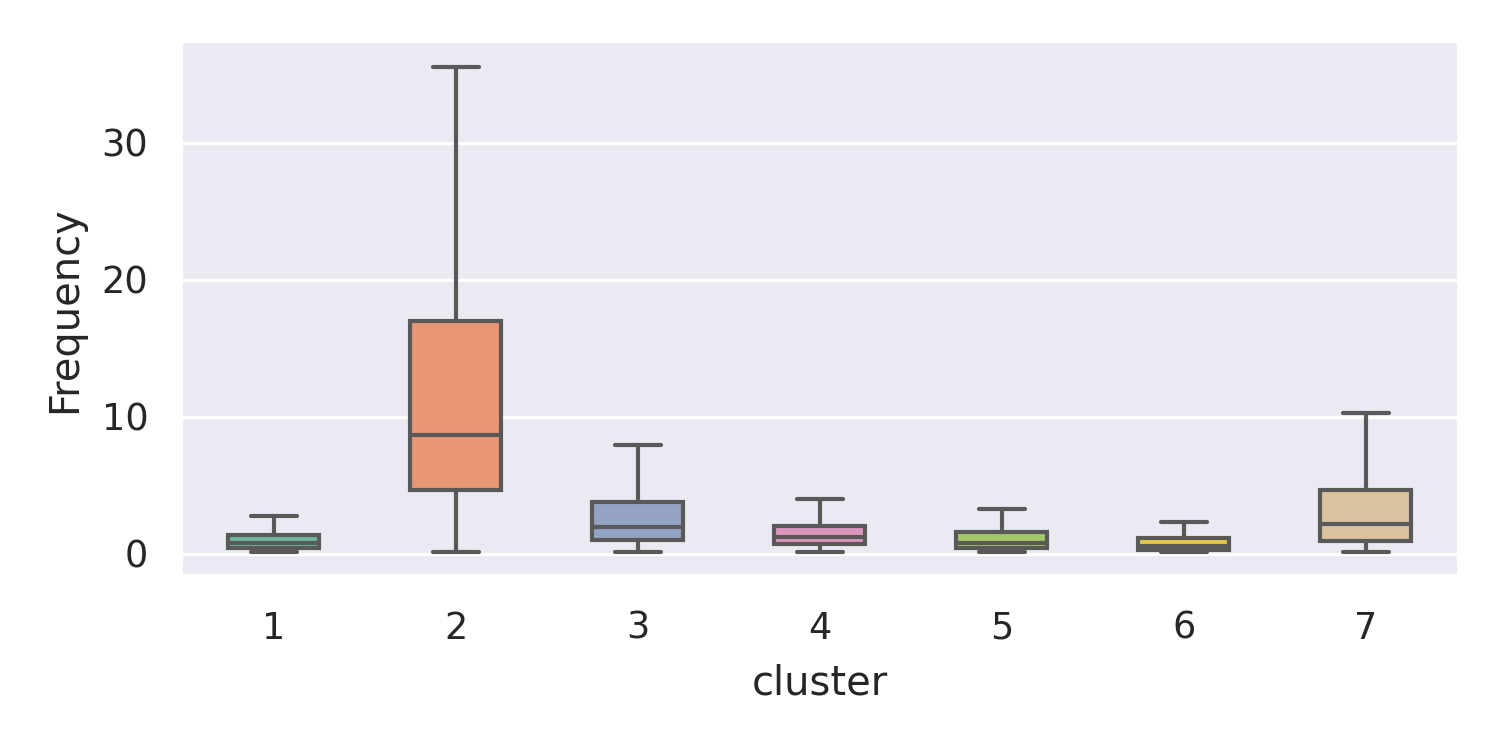}
    \caption{Box plot of trade frequency data for samples within each cluster. The median is represented by the middle line. The box encompasses the lower (25\%) and upper (75\%) quartiles. The whiskers, extending from the box, indicate the range from the minimum to the lower quartile and from the upper quartile to the maximum values.}
    \label{fig:freq}
\end{figure}

\paragraph{Size:} 
Let $\hat{q}_i$ indicate the size of $i$th order, the average \textbf{order size} is denoted by $\cfrac{1}{50}\sum\limits_{i=1}^{50}\hat{q}_i$. Additionally, we introduce another term called the average \textbf{trade size} $\cfrac{1}{50}\sum\limits_{i=1}^{50}q_i$ where $q_i$ is the filled quantity of $ith$ market order. It is worth noting that $\hat{q}_i$ and $q_i$ are not always the same, $\hat{q}_i$ represents the intended trade size, while $q_i$ represents the actual executed trade size, therefore $q_i \leq \hat{q}_i$. With the two terms, we will be able to construct another indicator \textbf{fill rate}, which is calculated as $\sum\limits_{i=1}^{50}q_i/(\sum\limits_{i=1}^{50}\hat{q}_i)$ (always $\leq 1$). 
\begin{figure}[h]
    \centering
    \includegraphics[width=0.5\linewidth]{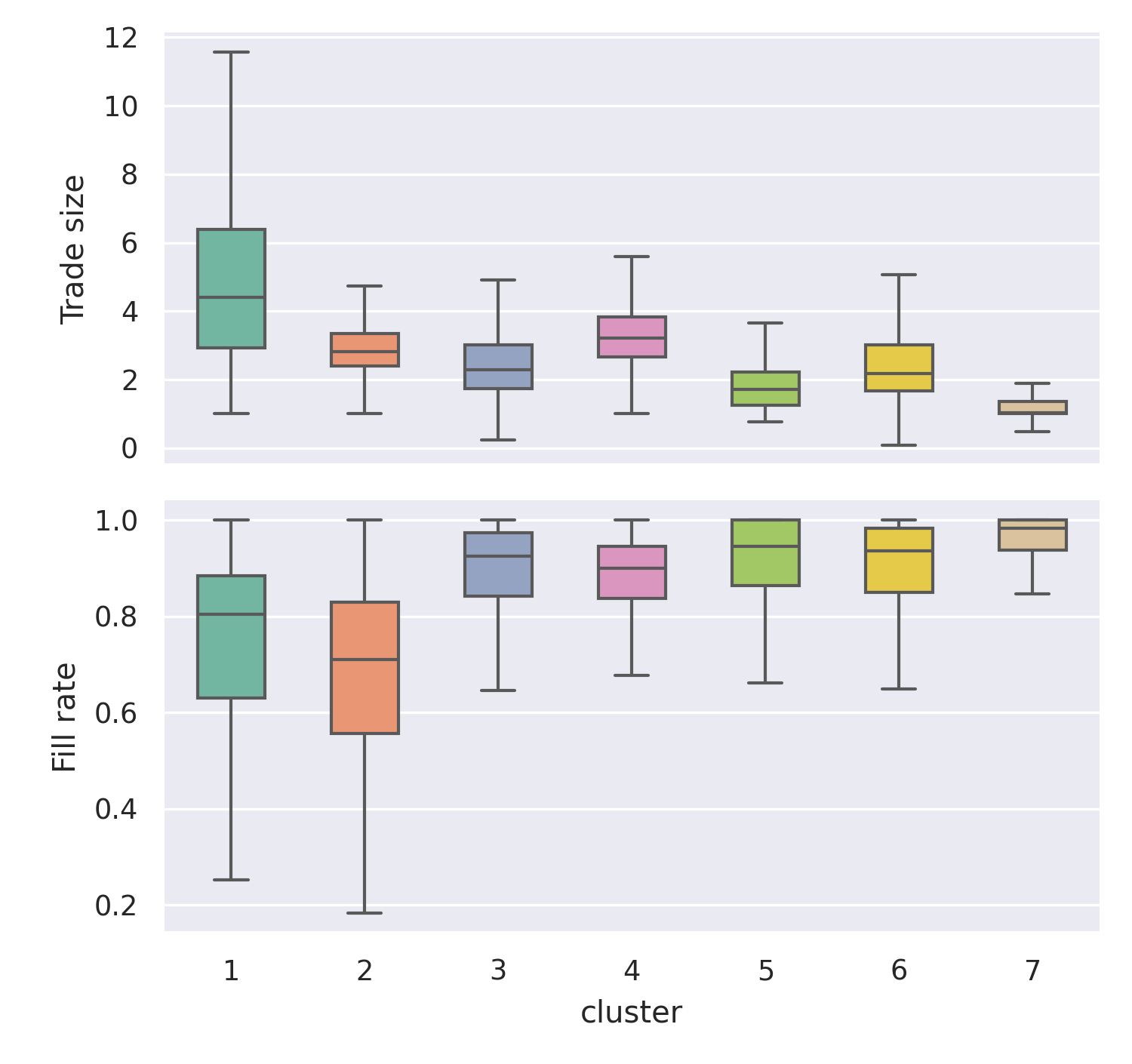}
    \caption{Box plot of trade size (top) and fill rate (bottom) data for samples within each cluster.}
    \label{fig:qty}
\end{figure}
\paragraph{Spread:} 
The average spread value is defined as $\cfrac{1}{50}\sum\limits_{i=1}^{50}(P^a_1(t_i-)-P^b_1(t_i-))$, where $P^a_1(t_i-)$(resp. $P^b_1(t_i-)$) is the best ask (resp. bid) price before the exceution of the $ith$ order. A low spread value indicates a more liquid market. 
\begin{figure}[h]
    \centering
    \includegraphics[width=0.5\linewidth]{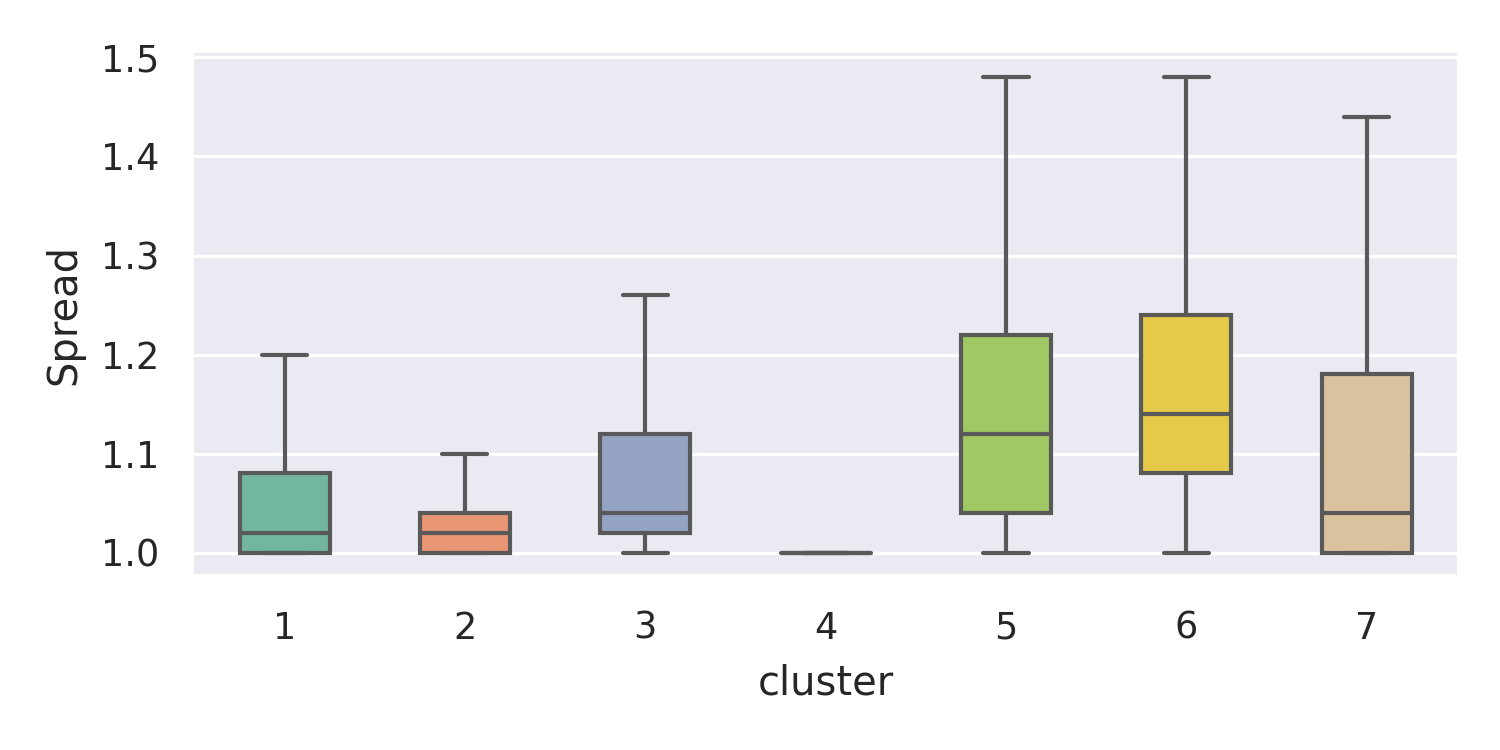}
    \caption{Box plot of bid-ask spread immediately before trades for samples within each cluster.}
    \label{fig:spread}
\end{figure}

\paragraph{Queue size (QS):} 
The indicator \textbf{queue size} is calculated as $\cfrac{1}{50}\sum\limits_{i=1}^{50}Q^{s_i}_1(t_i-)$. $s_i$ stands for the side of the $i$th order (buy or sell), and $Q^{s_i}_1(t_i-)$ denotes the volume of the available orders at the best level on the same side before the $i$th order was executed. Some traders may choose to place a market order when the available number of limit orders is low, in order to avoid missing out on potential gains. 

We also define another indicator (called \textbf{opposite queue size (opposite QS)}) as $\cfrac{1}{50}\sum\limits_{i=1}^{50}Q^{s_i^c}_1(t_i-)$. $s_i^c$ represents the opposite side of the $i$th order (buy or sell). $Q^{s_i^c}_1(t_i-)$ is the volume at the best level of the opposite side from where the trade occurs. For example, if order $i$ is a buy order $s_i=a$, $Q^{s_i^c}_1(t_i-)$ is the queue size at best bid limit.  A high value of \textbf{opposite queue size} implies that if the orders were placed as limit orders, they would take a long time to be executed due to the long waiting list. We may apply this indicator to measure the level of impatience displayed by an agent, which can serve as a valuable sign of aggressive actions by a market maker.

To gain a more comprehensive understand, we also analyzed the Related queue size (RQS) and the opposite related queue size (Opposite RQS), in addition to QS and opposite QS. RQS and opposite RQS are respectively defined by $\cfrac{1}{50}\sum\limits_{i=1}^{50}Q^{s_i}_1(t_i-)/q_i$ and $\cfrac{1}{50}\sum\limits_{i=1}^{50}Q^{s_i^c}_1(t_i-)/q_i$. A close to 1 RQS value indicates that the market orders clear almost the best level orders, hence the price moves after these trades.

We are inspired to include these indicators based on the observation that the arrival rate of order flows is influenced by the queue sizes. This property, named Queue Reactive, has been studied in several works \cite{huang2015simulating, wu2019queue}. 
\begin{figure}[h]
    \centering
    \includegraphics[width=0.8\linewidth]{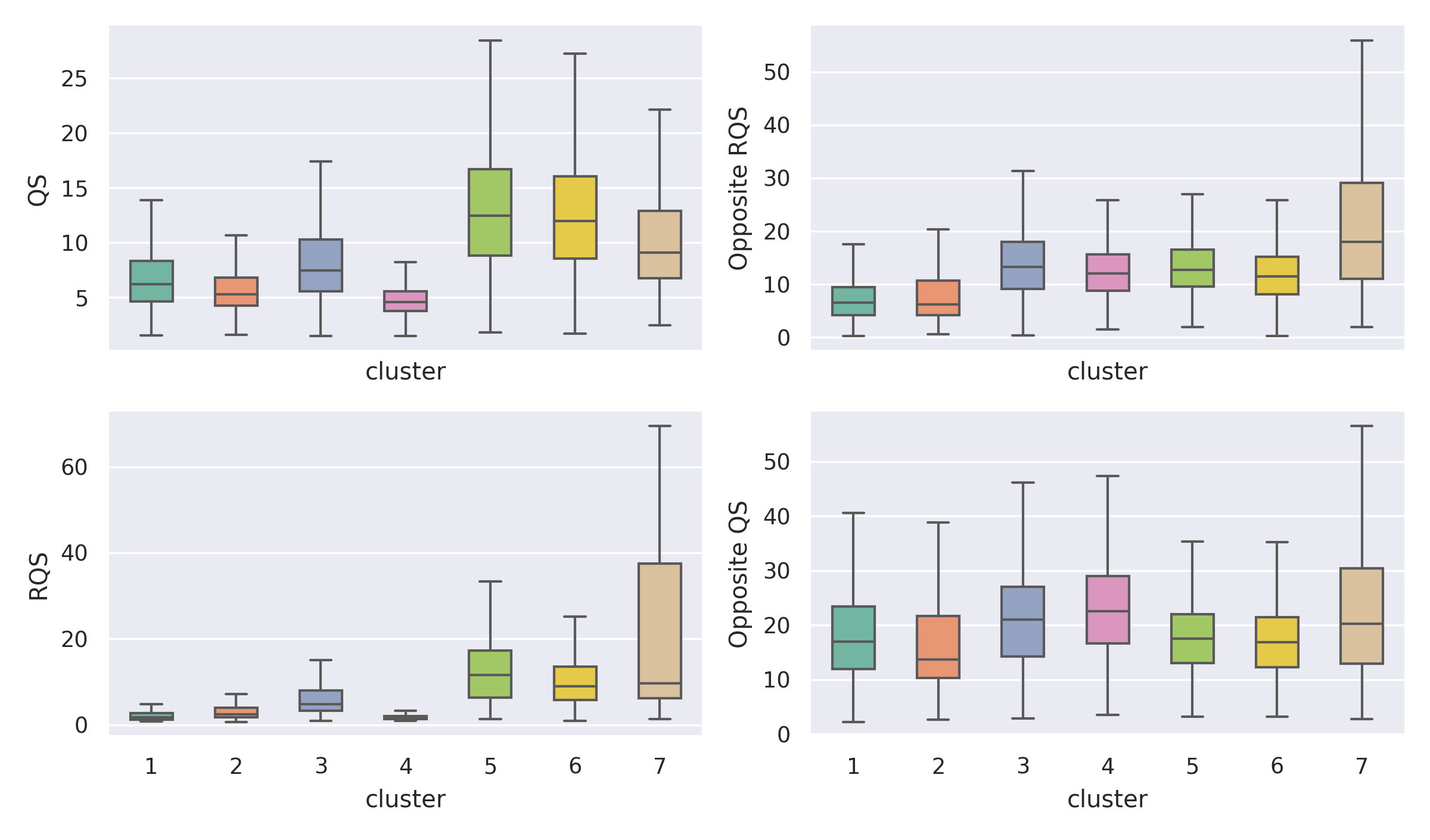}
    \caption{Within each cluster, box plot illustrate the queue sizes of samples. On the left side, the figures represent the queue sizes at the side where trades occur, while the right side figures stand for the queue sizes on the opposite side. The top figures show the actual queue sizes, while the bottom figures display the queue sizes relative to the trade size.}
    \label{fig:fear}
\end{figure}

\paragraph{Direction:} We represent the direction of an input using the formula $|\sum\limits_{i=1}^{50}q_i\cdot s_i|/(\sum\limits_{i=1}^{50}q_i)$, where $s_i$ is 1 if the order is a buy order, otherwise $s_i$ is -1. We take the absolute value in the formula because the crucial information we are interested in is whether the input is directional rather than the direction itself. A direction value close to 0 indicates a balanced input, while a value close to 1 indicates a highly directional input. Specifically, a value of 0 indicates that the buy and sell orders are evenly distributed, whereas a value of 1 indicates that all orders are on the same side. 
\begin{figure}[h]
    \centering
    \includegraphics[width=0.5\linewidth]{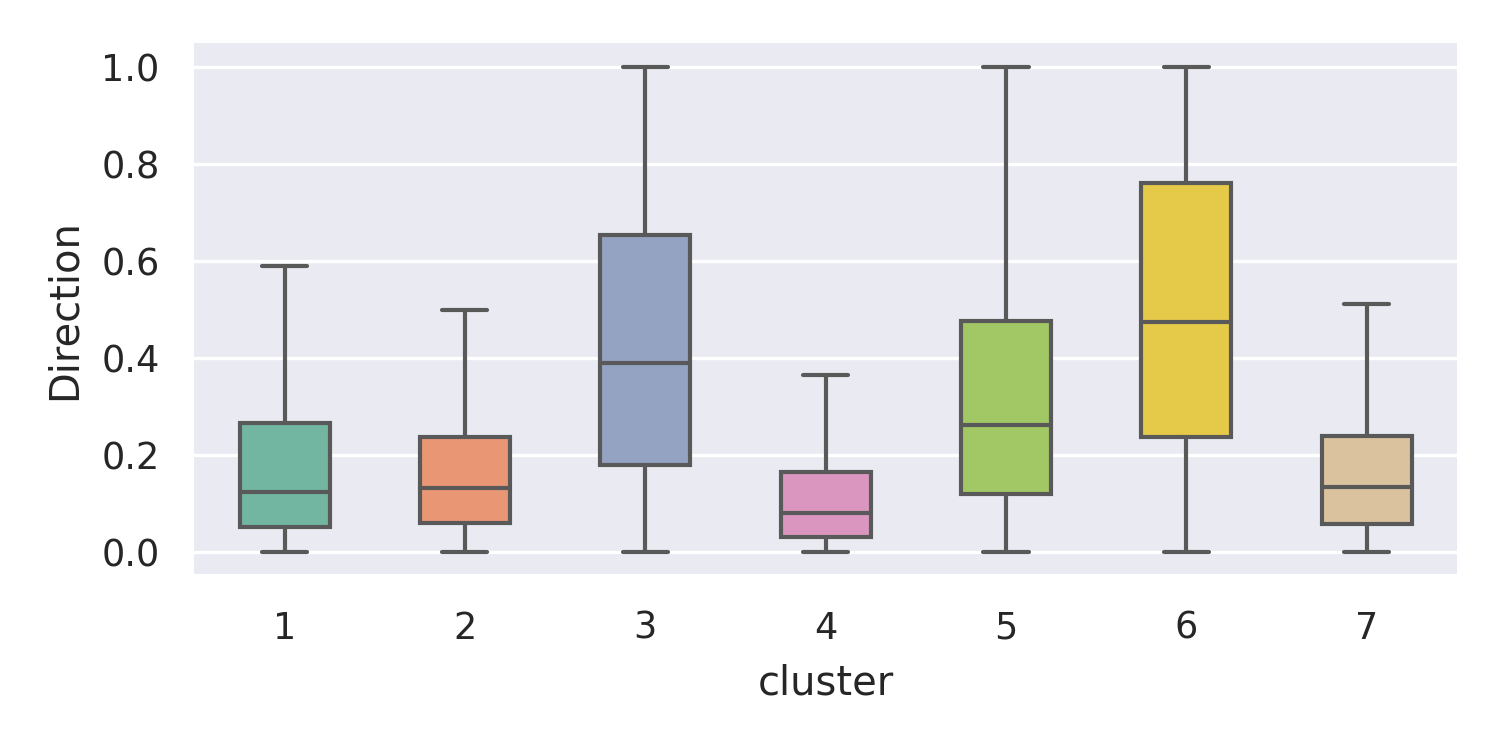}
    \caption{Box plots of direction indicators for samples within each cluster}
    \label{fig:direction}
\end{figure}

\paragraph{Limit to trade modification:} The final indicator we use is the proportion of modification in all the market orders, calculated as $\cfrac{1}{50}\sum\limits_{i=1}^{50} M_i$. Here $M_i$ indicates whether the $i$th market order is a limit to trade modification order. Let us note that $M=1$ means that the order was modified from an existing limit order to make it aggressive, while $M=0$ means that the order was added aggressive. 
\begin{figure}[h]
    \centering
    \includegraphics[width=0.5\linewidth]{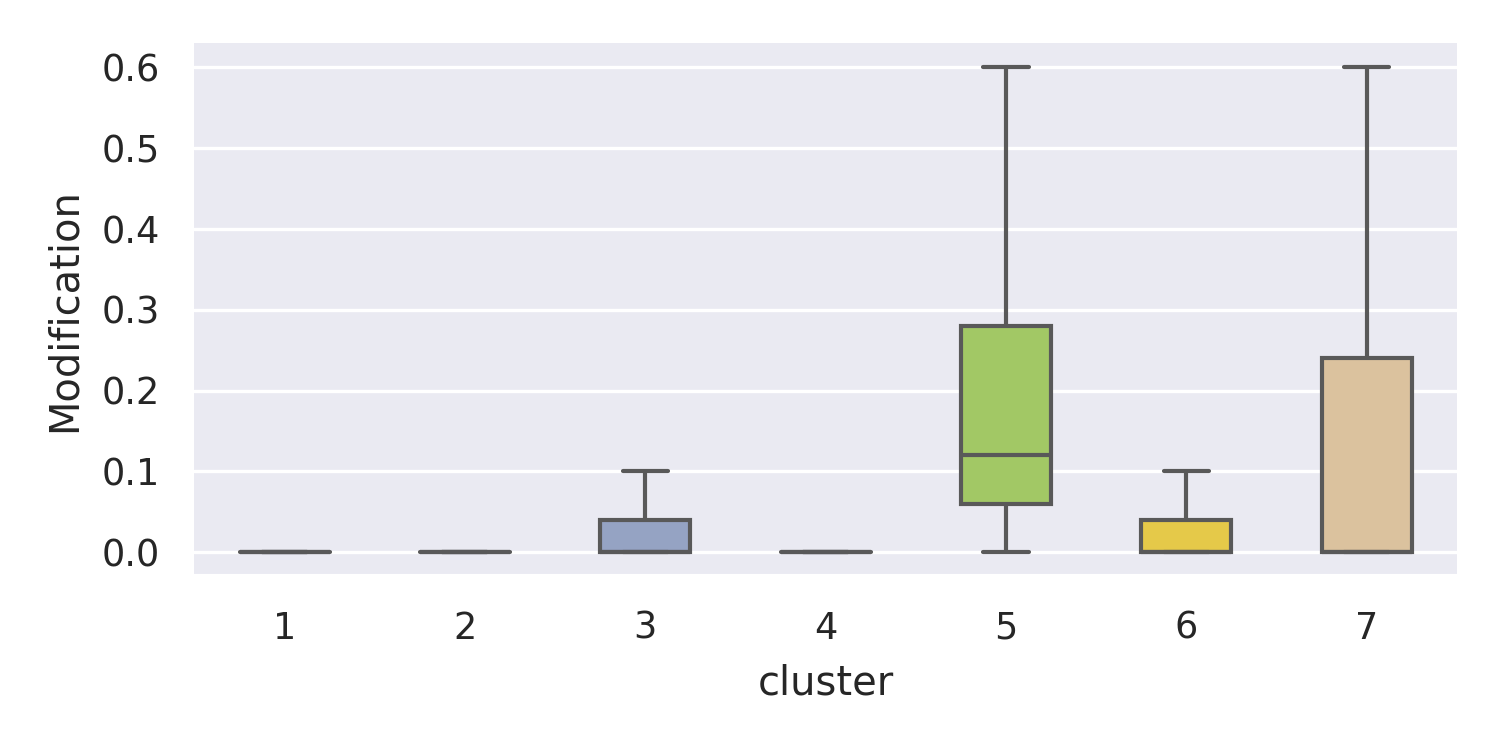}
    \caption{Box plots of modification proportion for samples within each cluster}
    \label{fig:limittotrade}
\end{figure}

Based on all previous indicator analysis, we can summarize the characteristics of these clusters in the following table \ref{tab:clust}, by using a rating system ranging from (+) to (+++).
\begin{table}[h]
    \centering
    \resizebox{\linewidth}{!}{
    \begin{tabular}{l|ccccccc}
        \hline
        cluster & 1 & 2 & 3 & 4 & 5 & 6 & 7 \\
        \hline
        Frequency & + & +++ & ++ &  + & + & + & ++ \\
        Trade size &  +++ & ++ & ++ & ++ & + & ++ & + \\
        Fill rate & + & + & ++ & ++ & ++ & ++ & +++ \\
        Spread & ++ & + & ++ & + & +++ & +++ & ++ \\
        QS & ++ & ++ & ++ & + & +++ & +++ & ++ \\
        Opposite QS & + & + & ++ & ++ & ++ & ++ & +++ \\
        Direction & + & + & +++ & + & ++ & +++ & + \\
        Modification &  &  & + & & ++ & +& ++ \\ 
        \hline
    \end{tabular}}
    \caption{Evaluation of the clusters based on the above indicators (from none() to low (+) to high (+++)}
    \label{tab:clust}
\end{table}

Notably, we see that 
\begin{itemize}
    \item Cluster 4 : This cluster exhibits low frequency, minimum spread, and zero modification. It is dominated by agents 10, 17, and 22. Referring to Figure \ref{fig:passive_ratio}, we observe that these three agents have almost no passive trades. Therefore in Cluster 4, agents primarily function as speculators.
    \item Cluster 6 : This cluster demonstrates low frequency, high bid-ask spread, and a significant directional indicator. Agents within this cluster perform directional trading.
    \item Cluster 7 : This cluster is characterized by a high opposite queue size, non-obvious direction and significant modification. Agent 11 and Agent 12 are the main contributors to this cluster. Examining Figure \ref{fig:passive_ratio}, we notice that these two agents exhibit a high passive-aggressive ratio, indicating that Cluster 7 represents the impatient behavior of market makers.
\end{itemize}

\subsection{Delving into details of each agent}
To further analyze the behavior of agents in different clusters, we use the indicators mentioned earlier to evaluate their samples in each cluster. We select a few agents as examples to illustrate the differences in their behavior across different clusters.

\begin{figure}[h]
\centering
    \begin{subfigure}[b]{\linewidth}
    \centering
    \includegraphics[width=0.8\linewidth]{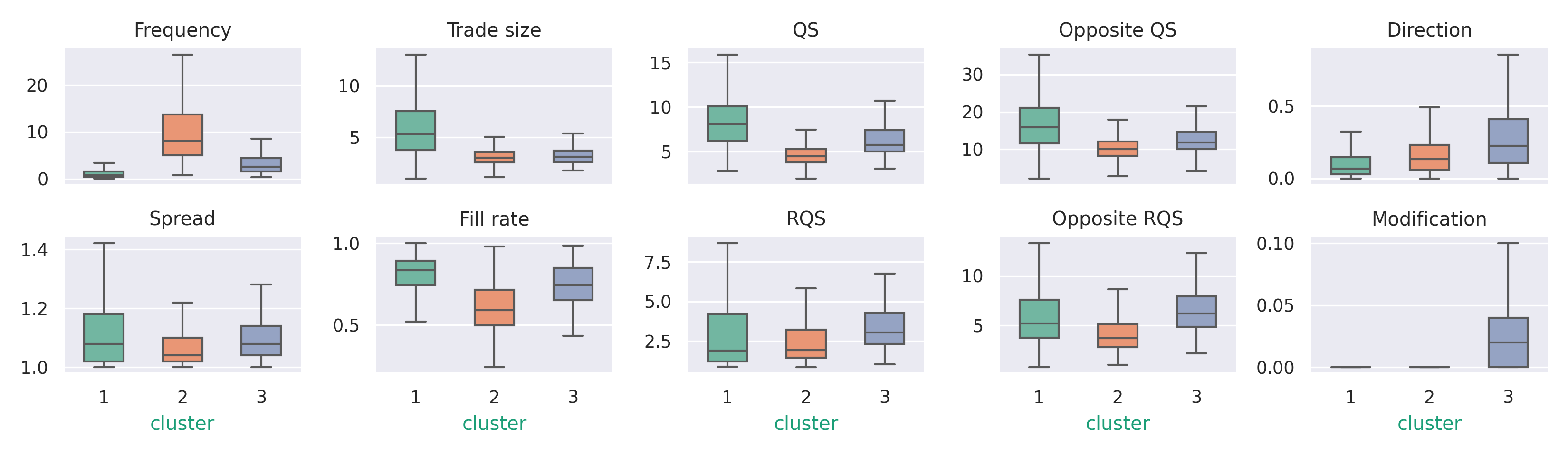}
    \caption{}
    \label{fig:ag9_box}
\end{subfigure}
\begin{subfigure}[b]{\linewidth}
    \centering
    \includegraphics[width=0.8\linewidth]{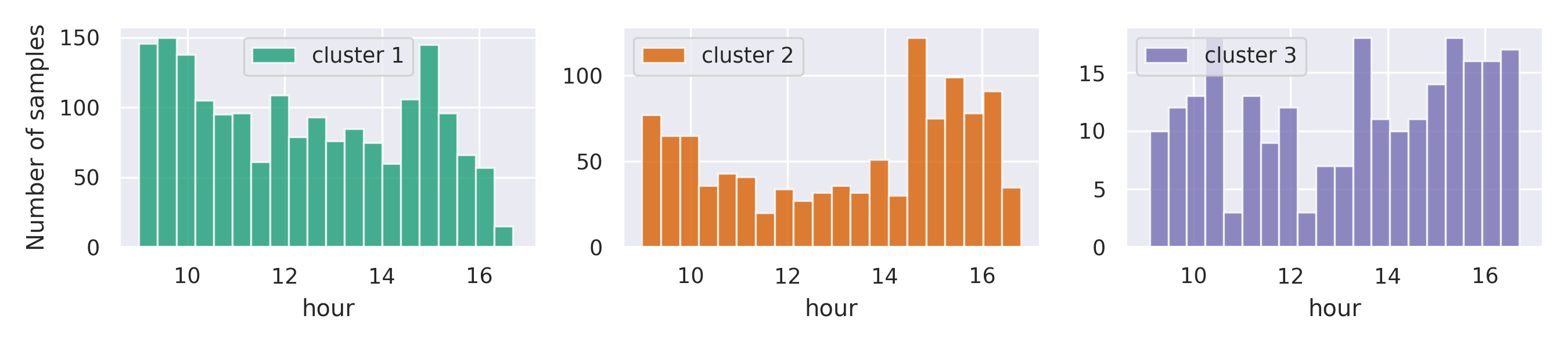}
    \caption{}
    \label{fig:ag9_hour}
\end{subfigure}
\caption{Agent 9. (a) Each figure corresponds to an indicator. Within each figure, the three vertical bars represent the performance of
samples from Agent 9 within each cluster. (b) Each figure corresponds to a cluster. Within each figure, the histogram plot  displays the distribution of samples selecting times.}
\label{fig:ag9}
\end{figure}

The first example is Member 9 which is, according to the results of Fig.\ref{fig:kmeans}, assigned to clusters 1,2 and 3. Figure \ref{fig:ag9_box} provides insight into its behavior within these clusters. In cluster 2, Agent 9 trades with a high-frequency way, and usually when the queue size is very low.
When we plot the time periods of these samples throughout the trading day (as shown in Figure \ref{fig:ag9_hour}), we observe that in the morning, Agent 9 preferentially behaves as Cluster 1, while in the afternoon, it exhibits behaviors similar to those of Cluster 2.

\begin{figure}[h]
    \centering
    \includegraphics[width=\linewidth]{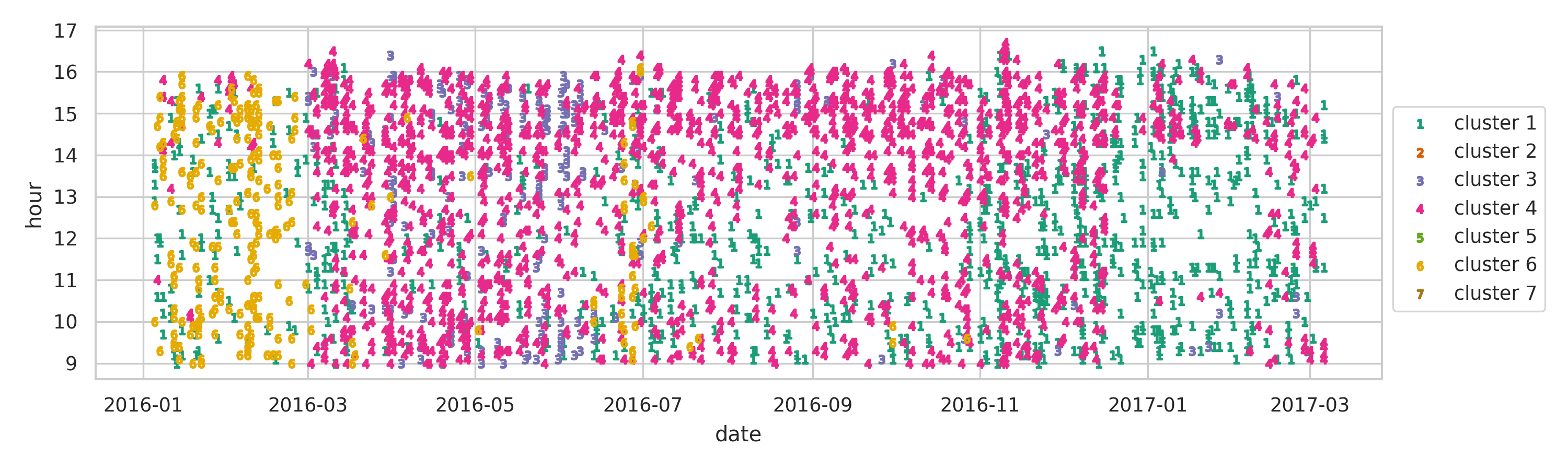}
\caption{2-D scatter plot. X-axis represents the dates and the y-axis represents the hour in a day. In this plot, each point stands for the occurring time of a selected sample and its color shows the cluster that it belongs to.}
\label{fig:ag10_hhdd}
\end{figure}

To visualize the evolution of agents' behaviors over time, we present two examples, namely agent 6 and 10, respectively in Figure \ref{fig:ag6_hhdd} and \ref{fig:ag10_hhdd}. During the period from January 2016 to March 2017, it is observed that Agent 10 significantly changes the behavior twice during this period. The first time of change occurs around March 2016, followed by another one around December 2016.
\subsection{Clusters visualization}
\begin{figure}[ht]
    \centering
    \includegraphics[width=0.9\linewidth]{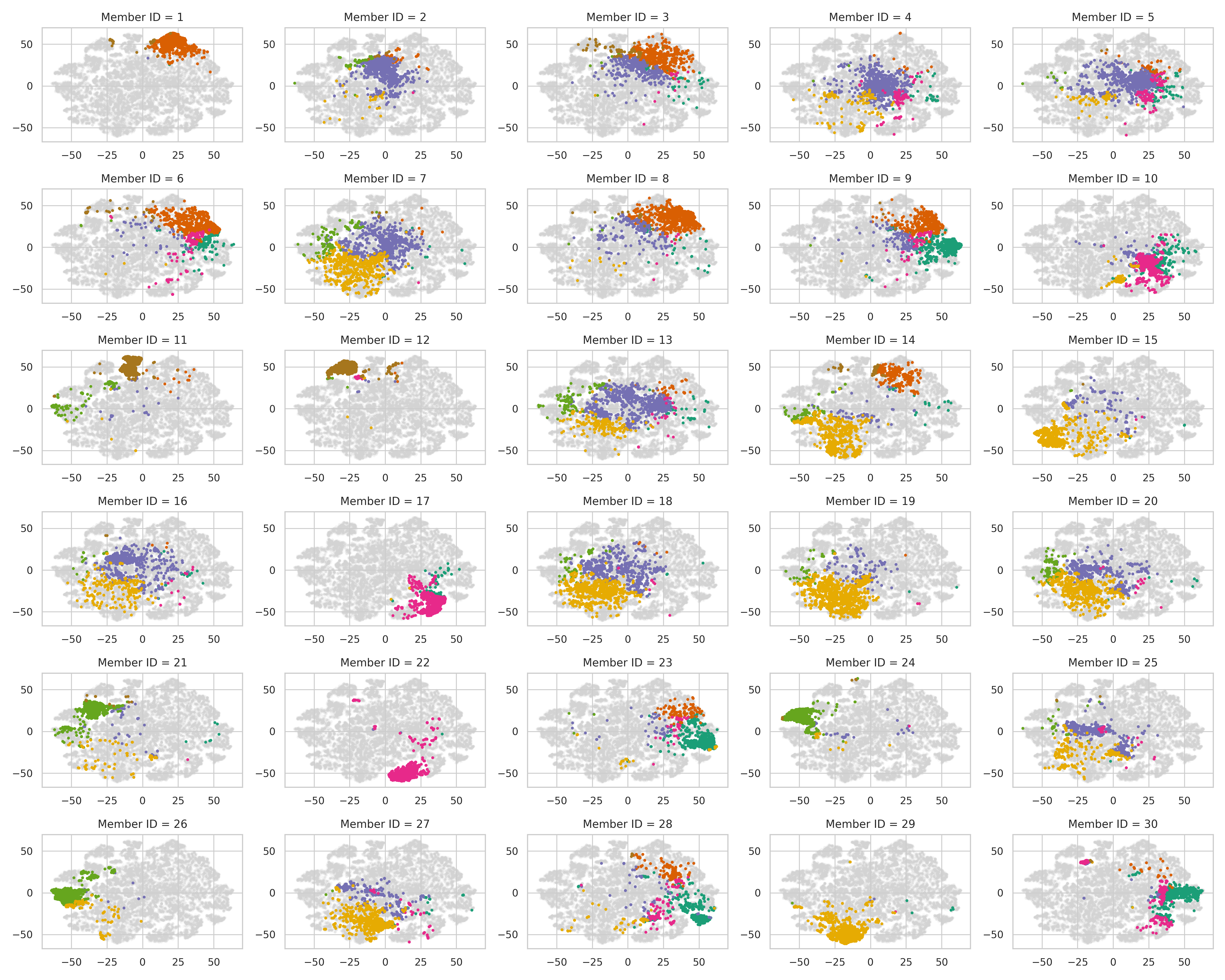}
    \caption{t-SNE visualization of 50,000 samples : the colored parts in each subfigure represent the samples of a given agent, the different colors indicate the clusters to which the agent belongs}
    \label{fig:tsne}
\end{figure}

A popular statistical method for visualizing hign-dimensional data is the t-distributed stochastic neighbor embedding (t-SNE) \cite{van2008visualizing}. It is a non-linear technique that maps high-dimensional data to a low-dimensional space while preserving the structure of the original data. However, in practice, t-SNE can be computationally expensive and struggle with high-dimensional data. Therefore, it is often recommended to first use another dimensionality reduction method, such as Principal component analysis (PCA), to reduce the number of dimensions to a reasonable amount before applying t-SNE. Figure \ref{fig:tsne} shows the results of applying t-SNE to 50,000 order book samples from the thirty agents.

Based on the t-SNE visualization, we have observed the following noteworthy observations:
\begin{itemize}
    \item Agent 11 and 12 are assigned to the same cluster, while their images are almost disjoint. This suggests that these two agents share similarities with respect to other agents in the dataset, but the difference between them is still distinct. Similar observations can be made for Agent 21, 24 and 26, although their dissimilarities are less clear compared to Agents 11 and 12.
    \item Even though Agent 1, 3, 6 and 8 all have a significant portion distributed within cluster 2, there is a higher degree of similarity among Agent 3,6 and 8 compared to the similarity between Agent 1 and the other three agents.
\end{itemize}
\section{Conclusion and Discussion}\label{sec:conclusion}

In this paper, we present a novel approach for limit order book analysis by designing a contrastive learning method with triplet loss. Our study uses the Cac40 index future data provided by Euronext Paris, spanning from January 2016 to February 2017. We make the assumption that individual agents maintain consistent behavior over short periods, while different agents exhibit distinct behaviors. By training neural networks, we obtain vector representations of sequences of market orders from the same agent.

We employ K-means clustering on the set of obtained representation vectors, in order to group the sequences of market orders effectively. This clustering cut the set to seven clusters. Subsequently, we define various indicators such as trading frequency, spread to characterize the sequences within each cluster. This allows us to identify distinct market marker clusters as well as clusters associated with directional agents. Furthermore, we analyze the behavior of each agent across different clusters based on these indicators, offering valuable insights into their trading behavior and its evolution over time.

In future research, we plan to expand our analysis to include both aggressive and passive trades, thus providing a more comprehensive understanding of the market. Inspired by the work \cite{cartea2023statistical}, we also intend to extend the order features by incorporating additional factors such as deep order volume in the limit order book and agent inventories.

Furthermore, the learned representation vectors can be applied to various downstream tasks. For instance, in market forecasting, the embedding vectors from active agents can effectively represent the market context. Additionally, these vectors can be utilized for agent-based generation of synthetic market data, offering new possibilities for market simulation and analysis.

\section*{Acknowledgements}
We thank Euronext Paris for making their data available to us. This research is partially supported by the Agence Nationale de la Recherche as part of the “Investissements d’avenir” program (reference ANR-19-P3IA-0001; PRAIRIE 3IA Institute).

\section*{Appendices}
\begin{appendices}
\section{More agent analysis}
\begin{figure}[h]
\centering
\begin{subfigure}[b]{\linewidth}
    \centering
    \includegraphics[width=0.8\linewidth]{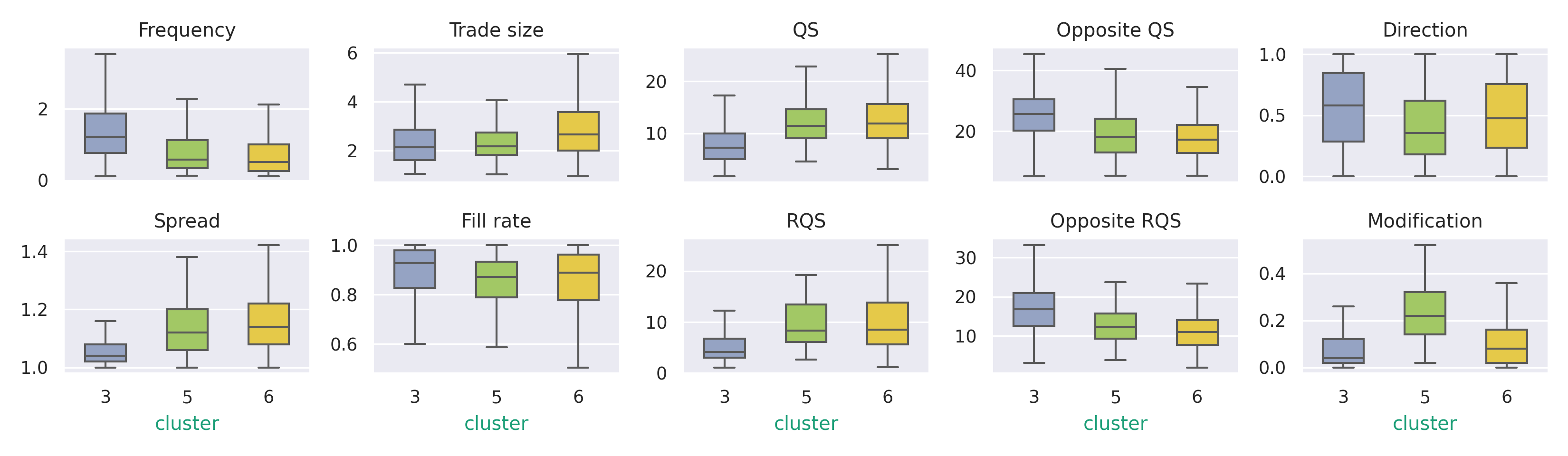}
    \label{fig:ag20_box}
    \caption{}
\end{subfigure}
\begin{subfigure}[b]{\linewidth}
    \centering
    \includegraphics[width=0.8\linewidth]{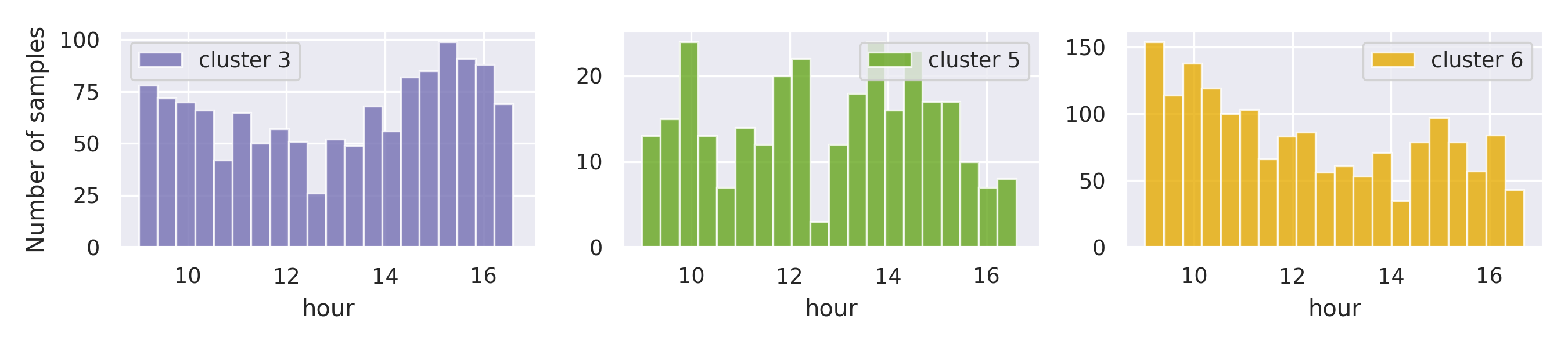}
    \label{fig:ag20_hour}
    \caption{}
\end{subfigure}    
\caption{Agent 20 : (a) Each figure corresponds to an indicator. Within each figure, the three vertical bars represent the performance of
samples from Agent 20 within each cluster. (b) Each figure corresponds to a cluster. Within each figure, the histogram plot  displays the distribution of samples selecting times.}
\label{fig:ag20}
\end{figure}

\paragraph{Member 20}\textbf{}\\
Agent 20, globally belongs to the clusters 3,5 and 6. We can observe that the samples of Agent 20 that belong to the cluster 3 have a higher frequency and are more likely to be active when the market is liquid as indicated by a smaller spread. In contrast, in cluster 5, with the significant modifications and lower direction index values, the agent behaves more like a market maker. (see Figure \ref{fig:ag20})
\paragraph{Member 6}\textbf{}\\
During the period from January 2016 to March 2017, it is observed that Agent 6 exhibits a decrease in activity while maintaining a relatively consistent behavior. 
\begin{figure}[h]
    \centering
    \includegraphics[width=\linewidth]{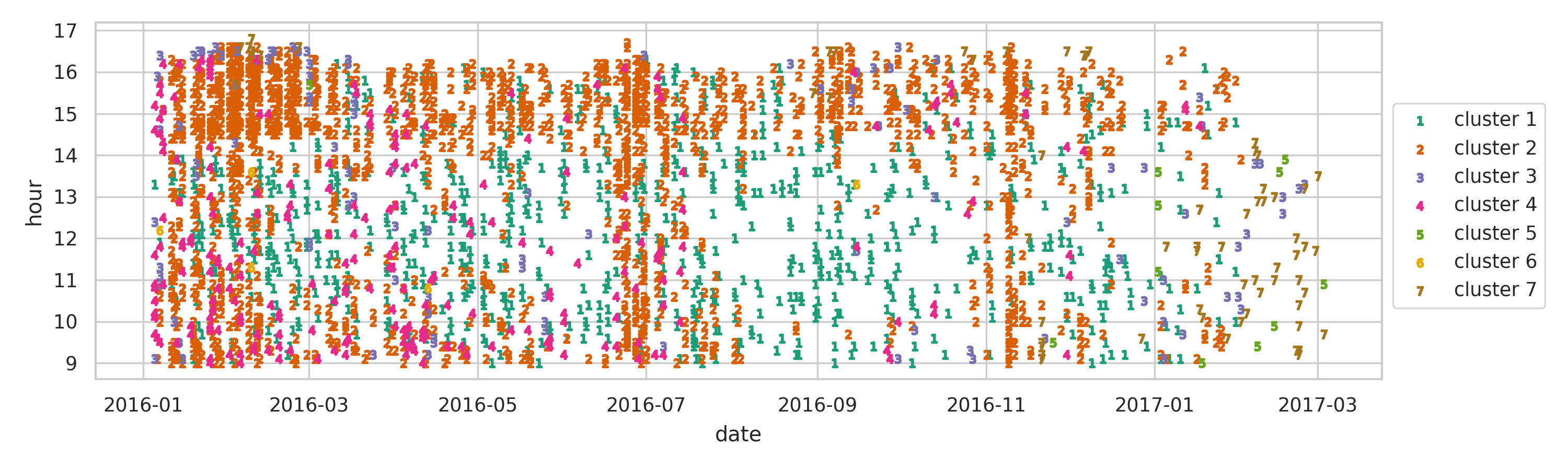}
    \caption{2-D scatter plot. X-axis represents the dates and the y-axis represents the hour in a day. In this plot, each point stands for the occurring time of a selected sample and its color shows the cluster that it belongs to.}
    \label{fig:ag6_hhdd}
\end{figure}

\end{appendices}

\begin{thebibliography}{10}

\bibitem{brogaard2010high}
Jonathan Brogaard et~al.
\newblock High frequency trading and its impact on market quality.
\newblock {\em Northwestern University Kellogg School of Management Working
  Paper}, 66, 2010.

\bibitem{brogaard2014high}
Jonathan Brogaard, Terrence Hendershott, and Ryan Riordan.
\newblock High-frequency trading and price discovery.
\newblock {\em The Review of Financial Studies}, 27(8):2267--2306, 2014.

\bibitem{cartea2023statistical}
{\'A}lvaro Cartea, Samuel~N Cohen, Rob Graumans, Saad Labyad, Leandro
  S{\'a}nchez-Betancourt, and Leon van Veldhuijzen.
\newblock Statistical predictions of trading strategies in electronic markets.
\newblock {\em Available at SSRN 4442770}, 2023.

\bibitem{chen2020simple}
Ting Chen, Simon Kornblith, Mohammad Norouzi, and Geoffrey Hinton.
\newblock A simple framework for contrastive learning of visual
  representations.
\newblock In {\em International conference on machine learning}, pages
  1597--1607. PMLR, 2020.

\bibitem{chopra2005learning}
Sumit Chopra, Raia Hadsell, and Yann LeCun.
\newblock Learning a similarity metric discriminatively, with application to
  face verification.
\newblock In {\em 2005 IEEE Computer Society Conference on Computer Vision and
  Pattern Recognition (CVPR'05)}, volume~1, pages 539--546. IEEE, 2005.

\bibitem{cont2023analysis}
Rama Cont, Mihai Cucuringu, Vacslav Glukhov, and Felix Prenzel.
\newblock Analysis and modeling of client order flow in limit order markets.
\newblock {\em Quantitative Finance}, pages 1--19, 2023.

\bibitem{gould2013limit}
Martin~D Gould, Mason~A Porter, Stacy Williams, Mark McDonald, Daniel~J Fenn,
  and Sam~D Howison.
\newblock Limit order books.
\newblock {\em Quantitative Finance}, 13(11):1709--1742, 2013.

\bibitem{grill2020bootstrap}
Jean-Bastien Grill, Florian Strub, Florent Altch{\'e}, Corentin Tallec, Pierre
  Richemond, Elena Buchatskaya, Carl Doersch, Bernardo Avila~Pires, Zhaohan
  Guo, Mohammad Gheshlaghi~Azar, et~al.
\newblock Bootstrap your own latent-a new approach to self-supervised learning.
\newblock {\em Advances in neural information processing systems},
  33:21271--21284, 2020.

\bibitem{gutmann2010noise}
Michael Gutmann and Aapo Hyv{\"a}rinen.
\newblock Noise-contrastive estimation: A new estimation principle for
  unnormalized statistical models.
\newblock In {\em Proceedings of the thirteenth international conference on
  artificial intelligence and statistics}, pages 297--304. JMLR Workshop and
  Conference Proceedings, 2010.

\bibitem{hagstromer2013diversity}
Bj{\"o}rn Hagstr{\"o}mer and Lars Nord{\'e}n.
\newblock The diversity of high-frequency traders.
\newblock {\em Journal of Financial Markets}, 16(4):741--770, 2013.

\bibitem{hastie2009elements}
Trevor Hastie, Robert Tibshirani, Jerome~H Friedman, and Jerome~H Friedman.
\newblock {\em The elements of statistical learning: data mining, inference,
  and prediction}, volume~2.
\newblock Springer, 2009.

\bibitem{he2020momentum}
Kaiming He, Haoqi Fan, Yuxin Wu, Saining Xie, and Ross Girshick.
\newblock Momentum contrast for unsupervised visual representation learning.
\newblock In {\em Proceedings of the IEEE/CVF conference on computer vision and
  pattern recognition}, pages 9729--9738, 2020.

\bibitem{hochreiter1997long}
Sepp Hochreiter and J{\"u}rgen Schmidhuber.
\newblock Long short-term memory.
\newblock {\em Neural computation}, 9(8):1735--1780, 1997.

\bibitem{hou2021stock}
Min Hou, Chang Xu, Yang Liu, Weiqing Liu, Jiang Bian, Le~Wu, Zhi Li, Enhong
  Chen, and Tie-Yan Liu.
\newblock Stock trend prediction with multi-granularity data: A contrastive
  learning approach with adaptive fusion.
\newblock In {\em Proceedings of the 30th ACM International Conference on
  Information \& Knowledge Management}, pages 700--709, 2021.

\bibitem{huang2015simulating}
Weibing Huang, Charles-Albert Lehalle, and Mathieu Rosenbaum.
\newblock Simulating and analyzing order book data: The queue-reactive model.
\newblock {\em Journal of the American Statistical Association},
  110(509):107--122, 2015.

\bibitem{Kingma2015}
Diederik~P. Kingma and Jimmy Ba.
\newblock Adam: A method for stochastic optimization.
\newblock {\em arXiv preprint arXiv:1412.6980}, 2015.

\bibitem{kirilenko2017flash}
Andrei Kirilenko, Albert~S Kyle, Mehrdad Samadi, and Tugkan Tuzun.
\newblock The flash crash: High-frequency trading in an electronic market.
\newblock {\em The Journal of Finance}, 72(3):967--998, 2017.

\bibitem{lloyd1982least}
Stuart Lloyd.
\newblock Least squares quantization in pcm.
\newblock {\em IEEE transactions on information theory}, 28(2):129--137, 1982.

\bibitem{macqueen1967classification}
J~MacQueen.
\newblock Classification and analysis of multivariate observations.
\newblock In {\em 5th Berkeley Symp. Math. Statist. Probability}, pages
  281--297, 1967.

\bibitem{mehari2022self}
Temesgen Mehari and Nils Strodthoff.
\newblock Self-supervised representation learning from 12-lead ecg data.
\newblock {\em Computers in biology and medicine}, 141:105114, 2022.

\bibitem{mohsenvand2020contrastive}
Mostafa~Neo Mohsenvand, Mohammad~Rasool Izadi, and Pattie Maes.
\newblock Contrastive representation learning for electroencephalogram
  classification.
\newblock In {\em Machine Learning for Health}, pages 238--253. PMLR, 2020.

\bibitem{oord2018representation}
Aaron van~den Oord, Yazhe Li, and Oriol Vinyals.
\newblock Representation learning with contrastive predictive coding.
\newblock {\em arXiv preprint arXiv:1807.03748}, 2018.

\bibitem{paszke2019pytorch}
Adam Paszke, Sam Gross, Francisco Massa, Adam Lerer, James Bradbury, Gregory
  Chanan, Trevor Killeen, Zeming Lin, Natalia Gimelshein, Luca Antiga, et~al.
\newblock Pytorch: An imperative style, high-performance deep learning library.
\newblock {\em Advances in neural information processing systems}, 32, 2019.

\bibitem{schroff2015facenet}
Florian Schroff, Dmitry Kalenichenko, and James Philbin.
\newblock Facenet: A unified embedding for face recognition and clustering.
\newblock In {\em Proceedings of the IEEE conference on computer vision and
  pattern recognition}, pages 815--823, 2015.

\bibitem{sirignano2019universal}
Justin Sirignano and Rama Cont.
\newblock Universal features of price formation in financial markets:
  perspectives from deep learning.
\newblock {\em Quantitative Finance}, 19(9):1449--1459, 2019.

\bibitem{sirignano2019deep}
Justin~A Sirignano.
\newblock Deep learning for limit order books.
\newblock {\em Quantitative Finance}, 19(4):549--570, 2019.

\bibitem{sohn2016improved}
Kihyuk Sohn.
\newblock Improved deep metric learning with multi-class n-pair loss objective.
\newblock {\em Advances in neural information processing systems}, 29, 2016.

\bibitem{sutskever2014sequence}
Ilya Sutskever, Oriol Vinyals, and Quoc~V Le.
\newblock Sequence to sequence learning with neural networks.
\newblock {\em Advances in neural information processing systems}, 27, 2014.

\bibitem{van2008visualizing}
Laurens Van~der Maaten and Geoffrey Hinton.
\newblock Visualizing data using t-sne.
\newblock {\em Journal of machine learning research}, 9(11), 2008.

\bibitem{wu2020conditional}
Hanwei Wu, Ather Gattami, and Markus Flierl.
\newblock Conditional mutual information-based contrastive loss for financial
  time series forecasting.
\newblock In {\em Proceedings of the First ACM International Conference on AI
  in Finance}, pages 1--7, 2020.

\bibitem{wu2019queue}
Peng Wu, Marcello Rambaldi, Jean-Fran{\c{c}}ois Muzy, and Emmanuel Bacry.
\newblock Queue-reactive hawkes models for the order flow.
\newblock {\em arXiv e-prints}, pages arXiv--1901, 2019.

\bibitem{zbontar2021barlow}
Jure Zbontar, Li~Jing, Ishan Misra, Yann LeCun, and St{\'e}phane Deny.
\newblock Barlow twins: Self-supervised learning via redundancy reduction.
\newblock In {\em International Conference on Machine Learning}, pages
  12310--12320. PMLR, 2021.

\bibitem{zhang2021deep}
Zihao Zhang, Bryan Lim, and Stefan Zohren.
\newblock Deep learning for market by order data.
\newblock {\em Applied Mathematical Finance}, 28(1):79--95, 2021.

\bibitem{zhang2019deeplob}
Zihao Zhang, Stefan Zohren, and Stephen Roberts.
\newblock Deeplob: Deep convolutional neural networks for limit order books.
\newblock {\em IEEE Transactions on Signal Processing}, 67(11):3001--3012,
  2019.

\end{thebibliography}
\end{document}